\documentclass[aps,preprint,showpacs,floatfix,superscriptaddress]{revtex4}
\usepackage{graphicx}

\begin{document}
\flushbottom

\title{Pressure dependence of the upper critical field of MgB$_2$ and of
YNi$_2$B$_2$C}
\author{H. Suderow}
\affiliation{Laboratorio de Bajas Temperaturas, Dpto F\'isica
Materia Condensada, Instituto de Ciencia de Materiales Nicol\'as
Cabrera, Universidad Aut\'onoma de Madrid, 28049 Madrid, Spain}
\author{V.G. Tissen}
\affiliation{Instituto de Ciencia de Materiales de Madrid, Consejo
Superior de Investigaciones Cient\'ificas, 28049 Madrid, Spain}
\author{J.P. Brison}
\affiliation{Centre des Recherches sur les Tr\`es Basses
Temp\'eratures CNRS, BP 166, 38042 Grenoble Cedex 9, France}
\author{J.L. Mart\'inez}
\affiliation{Instituto de Ciencia de Materiales de Madrid, Consejo
Superior de Investigaciones Cient\'ificas, 28049 Madrid, Spain}
\author{S. Vieira}
\affiliation{Laboratorio de Bajas Temperaturas, Dpto F\'isica
Materia Condensada, Instituto de Ciencia de Materiales Nicol\'as
Cabrera, Universidad Aut\'onoma de Madrid, 28049 Madrid, Spain}
\author{P. Lejay}
\affiliation{Centre des Recherches sur les Tr\`es Basses
Temp\'eratures CNRS, BP 166, 38042 Grenoble Cedex 9, France}
\author{S. Lee}
\affiliation{Superconductivity Research Laboratory, ISTEC,
1-10-13, Shinonome, Koto-ku, Tokyo, 135-0062 Japan}
\author{S. Tajima}
\affiliation{Superconductivity Research Laboratory, ISTEC,
1-10-13, Shinonome, Koto-ku, Tokyo, 135-0062 Japan}

\date{\today}

\begin{abstract}
We present measurements of H$_{c2}(T)$ under pressure in MgB$_2$
and in YNi$_2$B$_2$C. The changes in the shape of H$_{c2}(T)$ are
interpreted within current models and show the evolution of the
main Fermi surface velocities $v_F$ and electron-phonon coupling
parameters $\lambda$ with pressure. In MgB$_2$ the electron-phonon
coupling strength of the nearly two dimensional $\sigma$ band,
responsible for the high critical temperature, is more affected by
pressure than the $\pi$ band coupling, and the hole doping of the
$\sigma$ band decreases. In YNi$_2$B$_2$C, the peculiar positive
curvature of H$_{c2}(T)$ is weakened by pressure.
\end{abstract}

 \maketitle

\section{Introduction}

Recent advances in the synthesis and study of new materials have
shown that a highly non uniform electron-phonon coupling over the
Fermi surface can lead to many new, and unexpected,
superconducting properties\cite{Brandow03}. In MgB$_2$, which
superconducts at T$_c=40$ $K$\cite{Nagamatsu01}, the
electron-phonon coupling is very different in the two sets of
bands of the Fermi surface, the quasi two dimensional $\sigma$
bands, which strongly couple to the in-plane high energy E$_{2g}$
B-B phonon mode, and the three dimensional $\pi$ bands, leading to
the much discussed two band superconductivity
\cite{Liu01,An01,Choi02,Rubio01,Yelland02,Cubitt03}. Several new
effects are currently under discussion, as e.g. possible internal
modes and soliton structures \cite{Gurevich03,Tanaka02}. In the
non-magnetic nickel borocarbide superconductor YNi$_2$B$_2$C (with
T$_c=15.5K$ \cite{Canfieldgeneral}) the superconducting gap is
highly anisotropic and vanishes on points or lines of the Fermi
surface
\cite{Shulga,Yuan03,Yokoya00,Boaknin01,Izawa01,Martinez03}.
Superconducting electron tunnelling spectroscopy
\cite{Martinez03}, together with photoemission and point contact
experiments \cite{Yokoya00,Yanson97} demonstrated the importance
for the pairing interaction of a low energy (4meV) phonon mode,
softened by a nesting feature of the Fermi surface
\cite{Dugdale99}, a phenomenon previously observed in neutron
scattering experiments \cite{Zaretski99}.

The underlying physical properties of MgB$_2$ and YNi$_2$B$_2$C
are clearly different, and they represent interesting limiting
behaviors of electron-phonon coupled superconductors. MgB$_2$ is
the most clear-cut example of two band superconductivity, and
YNi$_2$B$_2$C has the highest reported anisotropy of the
superconducting gap \cite{Boaknin01}. It has been found that in
both compounds, the temperature dependence of the upper critical
field H$_{c2}(T)$ significantly deviates from the linear behavior
expected at low fields within isotropic, single-band, BCS s-wave
theory. Instead, the shape of H$_{c2}(T)$ shows a positive
curvature (PC), which is a direct consequence of the presence of
two well differentiated subgroups of electrons on the Fermi
surface. The Fermi velocities and electron-phonon coupling
parameters of these subgroups of electrons are obtained by fitting
the shape of H$_{c2}(T)$ within the effective two band model
described in Refs.\cite{Shulga,Dahm03} and in Appendix A.

Studying the changes induced by variations in the volume of the
unit cell on the different Fermi surface and electron-phonon
coupling parameters is clearly of paramount importance to get a
full understanding of this intriguing kind of superconductors, and
H$_{c2}(T)$ under pressure appears as a valuable tool to do so. As
a matter of fact, the pressure dependence of H$_{c2}(T)$ has been
previously used to obtain microscopic information of the Fermi
surface and coupling parameters of other kinds of superconducting
materials \cite{Brison}. However, although several studies address
H$_{c2}(T)$ at ambient pressure in MgB$_2$ and in
YNi$_2$B$_2$C\cite{Shulga,Dahm03,Kogan02,Lyard,Welp}, to the best
of our knowledge, no previous measurements of the evolution of
H$_{c2}(T)$ under pressure are available. Note that several groups
have systematically studied the effect of chemical pressure on the
upper critical field of MgB$_2$\cite{Kang04,Wilke04}. In that case
scattering effects dominate the observed behavior, leading to an
increase of the interband and/or intraband coupling, which smears
out the differences in the electronic properties and in the
coupling to the phonon modes between the subgroups of electrons
found in the pure compounds\cite{Golubov97}. The goal of this work
is to obtain information about the changes induced by pressure on
the different parts of the Fermi surface by measuring and
analyzing the pressure dependence of H$_{c2}(T)$ in MgB$_2$ and in
YNi$_2$B$_2$C.

\section{Experimental}

The single crystals of MgB$_2$ were grown as described in
\cite{LeeXtal}, and the ones of YNi$_2$B$_2$C in an image furnace
\cite{Martinez03}. Small samples of size 0.12x0.12x0.03 mm3, cut
from mother crystals, were loaded into a hole in NiMo-alloy gasket
positioned between two diamond anvils with 0.7 mm culet diameter.
The pressure in the cell was determined by the ruby fluorescence
method.

A susceptometer was designed to obtain the largest signal to noise
ratio, with a pick up coil wounded very close to the sample space.
Anvils were glued to the sapphire backing plates in order to
reduce the inductive coupling between coil systems used for ac
susceptibility measurements and the body of a diamond anvil cell
made of Cu-Be alloy. The secondary coils of two identical coil
systems were connected in opposite, thus forming a bridge. The
measuring coil system is mounted symmetrically around the anvils
and the reference system is placed nearby. After additional
compensation by means of an attenuator and a phase shifter the
total signal is detected with a lock-in amplifier.

Using this susceptometer, we measured the ac susceptibility as a
function of the temperature (down to 2 K) at different magnetic
fields (up to 9 T) for several fixed pressures in each compound.
The magnetic field was always applied perpendicular to the basal
plane of the crystallographic structure. The critical temperature
was determined the onset of the superconducting transition curves,
which was defined as the intersection of two tangents. One to the
flat portion of the curve above and the second to the steepest
variation in the signal below superconducting transition, as shown
by the lines in Fig.1a and b. In these figures we show a
representative example of the temperature dependence of the
susceptibility around several superconducting transitions,
measured at different magnetic fields and at 13.4 GPa in MgB$_2$,
and 2.3 GPa in YNi$_2$B$_2$C. From the dependence of the critical
temperature as a function of the magnetic field at each pressure,
we obtain the data shown in figures 2 and 4 and discussed in the
rest of the paper.

On the other hand, we use a methanol-ethanol mixture as a
pressure-transmitting medium, which is thought to give
quasi-hydrostatic pressure conditions. It is important to
emphasize that previous measurements of T$_c(P)$ at zero magnetic
field in hydrostatic conditions (with helium as a pressure
transmitting medium, see e.g.\cite{Deemyad,Goncharov02} and
references therein), made in MgB$_2$, agree well with the data
presented here. Hence, we can exclude that the eventual deviations
from hydrostatic conditions influence the results obtained here.

\section{Results and discussion}

To interpret our data, we follow the microscopic model of
Refs.\cite{Shulga,Dahm03}, which numerically linearizes the
equations for H$_{c2}(T)$ obtained from Eliashberg theory
\cite{Prohammer87} (see Appendix A). The main change of
H$_{c2}(T)$ under pressure occurs in the form of the PC and the
value of H$_{c2}(T=0$ $K)$, which are uniquely determined by the
values of coupling strengths, $\lambda_1,\lambda_2$ and Fermi
velocities $v_{F1}$ and $v_{F2}$ of the different subgroups of
electrons used to model the Fermi surface\cite{Shulga}. Here, we
take for the interband coupling constants
$\lambda_{12}=\lambda_{21}=\lambda_{2}$. In fact, the values found
in literature $\lambda_{12}$ and $\lambda_{21}$ are indeed close
to $\lambda_{2}$ \cite{Shulga,Dahm03}, and they show the same
pressure dependencies, if left as free parameters, as
$\lambda_{2}$. Other parameters of the model, the Coulomb
pseudopotential and a mean phonon frequency do not affect the PC
and form of $H_{c2}(T)$, and we therefore leave them unchanged
(see Table I).

\subsection{Upper critical field under pressure of MgB$_2$.}

The temperature dependence of the upper critical field of MgB$_2$
up to 20.5 GPa is shown in Fig.\ 2. We could not find any previous
H$_{c2}(T,P)$ data under pressure, although, as already noted,
T$_c(P,H=0)$ and H$_{c2}(T,P=0)$ have been thoroughly studied in
Refs.\cite{Kogan02,Lyard,Welp,Deemyad,Goncharov01,Goncharov02,Tomita01},
and agree well with our data. More recently, a broadening of the
transition under magnetic fields has been discussed in terms of
fluctuation in Ref.\cite{Masui}. However, it appears to depend on
the volume of the samples.

The critical temperature drops by a factor of 2 at 20.5 GPa, and
H$_{c2}(T=0$ $K)$ by a factor of 4. The shape of H$_{c2}(T)$
changes by increasing pressure. The PC is always observed,
although it appears to be slightly weakened by pressure. More
detailed information is found by fitting the data (lines in Fig.\
2) and following the evolution of the most important parameters,
as shown in Fig.\ 3. The Fermi velocities (Fig.\ 3a) decrease for
both sets of bands under pressure. The slightly stronger decrease
of $v_{F\sigma}$ reflects the stronger pressure dependence of the
band structure of this subgroup of electrons. The dispersion
relation can be taken to be roughly quadratical in the basal
plane, as the Fermi level cuts these bands near their top
\cite{An01,Yelland02}, so that the Fermi velocity $v_{F\sigma}$ is
approximately proportional to the radius of the $\sigma$ sheets.
Hence, the decrease observed here shows the continuous shrinking
of the volume of the $\sigma$ sheet under pressure. As emphasized
in early band structure calculations on MgB$_2$ and related
isoelectronic systems\cite{An01}, the layer of Mg$^{2+}$ ions
causes charge transfer from the $\sigma$ to the $\pi$ bands,
doping with holes the $\sigma$ bands, which strongly contribute to
the density of states at the Fermi level due to their
two-dimensionality. Our data show how pressure reduces the doping,
and the $\sigma$ band density of states. This reduction appears to
affect mainly the electron-phonon coupling in the $\sigma$ bands
(Fig.\  3b). As a matter of fact, previous estimations of the
pressure effect on the overall coupling constant $\lambda$, based
on the T$_c(P)$ measurements\cite{Deemyad,Goncharov01}), agree
with the pressure dependence of $\lambda_{\sigma}$ found here.
Hence, the overall decrease in the electron-phonon coupling under
pressure does not occur over the whole Fermi surface, but must be
associated to a loss of electron-phonon coupling mainly in the
$\sigma$ bands, which is ultimately responsible for the decrease
of T$_c$(P).

On the other hand, we should also emphasize that the high energy
E$_{2g}$ phonon mode, to which the $\sigma$ band electrons mainly
couple, has been shown by Raman scattering measurements to stiffen
under pressure, increasing its frequency by about 30\% at 20
GPa\cite{Goncharov01,Goncharov02}. Correspondingly,
$\lambda_{\sigma}$ decreases, as shown here in Fig.\  3b, because
the hole doping of the $\sigma$ bands, achieved through the ionic
layered character of MgB$_2$, is gradually lost. Note also that
previous experiments have identified a kink in T$_c(P)$ at 20 GPa,
which has been attributed to an electronic topological transition,
corresponding to one of the two $\sigma$ sheets moving below the
Fermi level at 20 GPa \cite{Meletov02,Deemyad,Goncharov01}. The
decrease of $v_{F\sigma}$ observed here shows the pressure induced
filling of the $\sigma$ bands, and agrees well with this scenario.

\subsection{Upper critical field under pressure of YNi$_2$B$_2$C.}

The pressure dependence of the critical temperature and upper
critical field of YNi$_2$B$_2$C is shown in Fig.\  4. Previous
T$_c$(P) measurements were done at zero field\cite{Alleno95}, and
stopped at a much lower pressure (3 GPa), so that the decrease
measured here most clearly above 3 GPa (inset of Fig.\  4) could
not be observed. H$_{c2}$(T) at ambient pressure is in excellent
agreement with previous work, and shows the extreme PC discussed
in Ref.\cite{Shulga}.  Remarkably, while the critical temperature
is suppressed by a factor of 2 at 10 GPa, the zero temperature
extrapolation of the upper critical field drops by an order of
magnitude. In addition, the PC in H$_{c2}(T)$ is also suppressed
by pressure. The best fit to the data is shown as lines in Fig.\
4. At ambient pressure, we obtain the parameters summarized in
Table I. Their evolution as a function of pressure is shown in
Fig.\ 5. Note that the electron-phonon coupling parameters
$\lambda_1$ and $\lambda_2$ drop both continuously with pressure
(Fig.\ 5b) in approximately the same way. However, the changes in
the Fermi velocity occur mainly in the subgroup of electrons with
strongest coupling ($v_{F1}$, Fig.\ 5a), evidencing an increase of
the volume of the corresponding part of the Fermi surface.

At zero pressure, the Fermi surface nesting feature found in
Ref.\cite{Dugdale99} leads to a significant softening of an
acoustic phonon mode down to about 4meV. The corresponding
strong-coupling feature has been observed in tunnelling
spectroscopy, and an electron-phonon coupling parameter $\lambda$
between $0.5$ and $0.8$, of the same order as the one used here
for $\lambda_1$ at ambient pressure (Table I), has been
obtained\cite{Martinez03}. The decrease in the electron-phonon
coupling under pressure shown in Fig.\ 5b should be related to the
hardening of this mode and the concomitant disappearance of the
Fermi surface nesting feature. This must be accompanied by an
increase of the volume of the corresponding part of the Fermi
surface, explaining the increase of $v_{F1}$ observed here (Fig.\
5a). The importance of band structure effects in understanding the
microscopic origin for the peculiar superconducting properties of
YNi$_2$B$_2$C is evidenced by our data.

\section{Summary and conclusion}

In summary, we have found that the evolution of H$_{c2}$(T) under
pressure in MgB$_2$ and in YNi$_2$B$_2$C reflects the changes
occurring in the bandstructure and electron-phonon coupling of the
different subgroups of electrons on the Fermi surface. In the two
band superconductor MgB$_2$, the electron-phonon coupling in the
$\sigma$ band, responsible for its high critical temperature, is
strongly affected by pressure through the reduction of the
$\sigma$ band hole doping. In YNi$_2$B$_2$C, the critical
temperature decreases continuously with pressure, together with
the electron-phonon coupling parameters in both subgroups of
electrons. H$_{c2}$(T = 0 K) dramatically decreases under
pressure, and, at the same time, the peculiar PC curvature of
H$_{c2}$(T) is weakened, showing the pressure induced increase in
the volume of the Fermi surface part corresponding to the strong
coupling subgroup of electrons.

\section{Acknowledgments}

We specially acknowledge discussions and help of P.C. Canfield. We
are also grateful to F. Guinea and A. Levanyuk and for receiving
support from the ESF programme VORTEX, from the MCyT (Spain;
grants MAT-2001-1281-C02-0, MAT-2002-1329 and SAB2000-039), and
from the Comunidad Aut\'onoma de Madrid (07N/0053/2002, Spain).
This work was partially supported by the New Energy and Industrial
Technology Development Organization (NEDO) as the Collaborative
Research and Development of Fundamental Technologies for
Superconductivity Applications. The Laboratorio de Bajas
Temperaturas is associated to the ICMM of the CSIC.

\newpage
\appendix
\section{}

Equations for two band superconductors have been given by numerous
authors \cite{Prohammer87,Shulga}. For the sake of completeness,
we give here the linearized gap equations under magnetic field,
for a superconductor in the clean limit (negligible impurity
scattering), as can be found in reference \cite{Shulga} (with
corrections of typewriting mistakes of factor 2 in the Matsubara
frequency and definition of $\lambda_{i,j}(n)$). In the following,
i and j are the band index, $\alpha_{i,j}F(\omega)$ is the density
of interactions from band i to j (involving diffusion from band i
to j), $v_{Fi}$ is the Fermi velocity of band i,  and $H_{c2}$ is
the upper critical field:

\begin{eqnarray}
\tilde{\Delta}_i(n)& = & \pi T\sum_{j,m}\left[\lambda_{i,j}(m-n)-\mu^*\delta_{i,j}\theta(\omega_c-\mid \omega_m \mid)\right] \nonumber\\
&& \qquad \times \chi_j(m)\tilde{\Delta}_j(m)(\omega_m) \label{equationgap}\\
&&\nonumber\\
\tilde{\omega}_i(n) & = & \omega_n + \pi T\sum_{j,m}[\lambda_{i,j}(m-n)]sgn(\omega_m)\nonumber\\
&&\nonumber\\
\chi_i(n) & = & (2/\sqrt{\beta_i}) \int_{0}^{\infty }dq exp(-q^2) \nonumber\\
&&\times tan^{-1} \left\{ \frac{ q\sqrt{\beta_i} }{\left[ \mid
\tilde{\omega}_i(n)\mid + i\mu_B H_{c2} sgn(\omega_n)\right]}
\right\}
\nonumber%\label{equationchi}
\end {eqnarray}

\begin {eqnarray*}
with \quad \beta_i &=& \pi H_{c2}v_{Fi}^2/(2\phi_0),   \hspace{1cm}\omega_n   =  \pi T(2n+1) \\
\lambda_{i,j}(n-m) &=& 2\int_{0}^{\infty } d\omega
\alpha_{i,j}^2\omega \frac{F(\omega)}{\omega^2 +
(\omega_n-\omega_m)^2}
\end {eqnarray*}

The value of $H_{c2}$ is given by the largest set of values of
$\beta_i$ yielding a non trivial solution of equation
\ref{equationgap}.

We have used in addition an Einstein spectrum for the density of
interactions:
\begin{equation}
\alpha_{i,j}(\omega)F(\omega)=\frac{\lambda_{i,j}\Omega}{2}\delta(\omega-\Omega)
\end{equation}
introducing a characteristic energy scale $\Omega$. In the
calculations, we find $\Omega\approx21K$: let us note that this
value of $\Omega$ cannot be determined by the fit of $H_{c2}$
only: it depends directly of the absolute value of the
$\lambda_{i,j}$ whereas the shape of $H_{c2}$ is controlled only
by the relative size of the $\lambda_{i,j}$ (as long as they are
not too large). The cut-off parameter $\omega_c$ in equation
\ref{equationgap} is taken to $10\Omega$, and the infinite system
of equations \ref{equationgap} has been truncated at each
temperature when $\mid\omega_n\mid$ reaches $\omega_c$.

Note also that bare Fermi velocities enter the above equations,
and that renormalized Fermi velocities
$v_{Fi}^*=v_{Fi}/(1+\sum_{j}\lambda_{i,j})$ should be used for
comparison to specific heat or de Haas-van Alphen measurements
\cite{Mazin2003}.

\newpage

\begin{table}{
\begin{tabular}{c|c|c|c|c|c|}
    & $\lambda_1$ & $\lambda_2$ & v$_{F1}$ & v$_{F2}$ \\\hline
  MgB$_2$ & 1.1 & 0.35 & 0.29 $10^6m/s$ & 0.9 $10^6m/s$ \\ \hline
  YNi$_2$B$_2$C & 0.84 & 0.27 & 0.059 $10^6m/s$ & 0.7 $10^6m/s$ \\
\end{tabular}
}\caption{The table shows the parameters used for the ambient
pressure fits. We have taken $\mu^*=0.1$\protect, and a mean
phonon frequency of $\theta=404K$ for MgB$_2$ and $\theta=248K$
for YNi$_2$B$_2$C. All parameters are similar to the ones used in
previous work \protect\cite{Shulga,Dahm03}. Slight differences
arise from a simpler choice of the matrix of strong coupling
parameters (only two different values ) describing the coupling
anisotropy.  In the case of YNi$_2$B$_2$C, our parameters agree
well with the tunnelling spectroscopy experiments of
\protect\cite{Martinez03}. For MgB$_2$,
$\lambda_{\sigma}=\lambda_1, \lambda_{\pi}=\lambda_2=
\lambda_{\pi\sigma}=\lambda_{12}=\lambda_{\sigma\pi}=
\lambda_{21}$ and $v_{\sigma}=v_{F1}, v_{\pi}=v_{F2}$.}
\end{table}

\newpage

\begin{figure}[ht]
\includegraphics[width=13cm,clip]{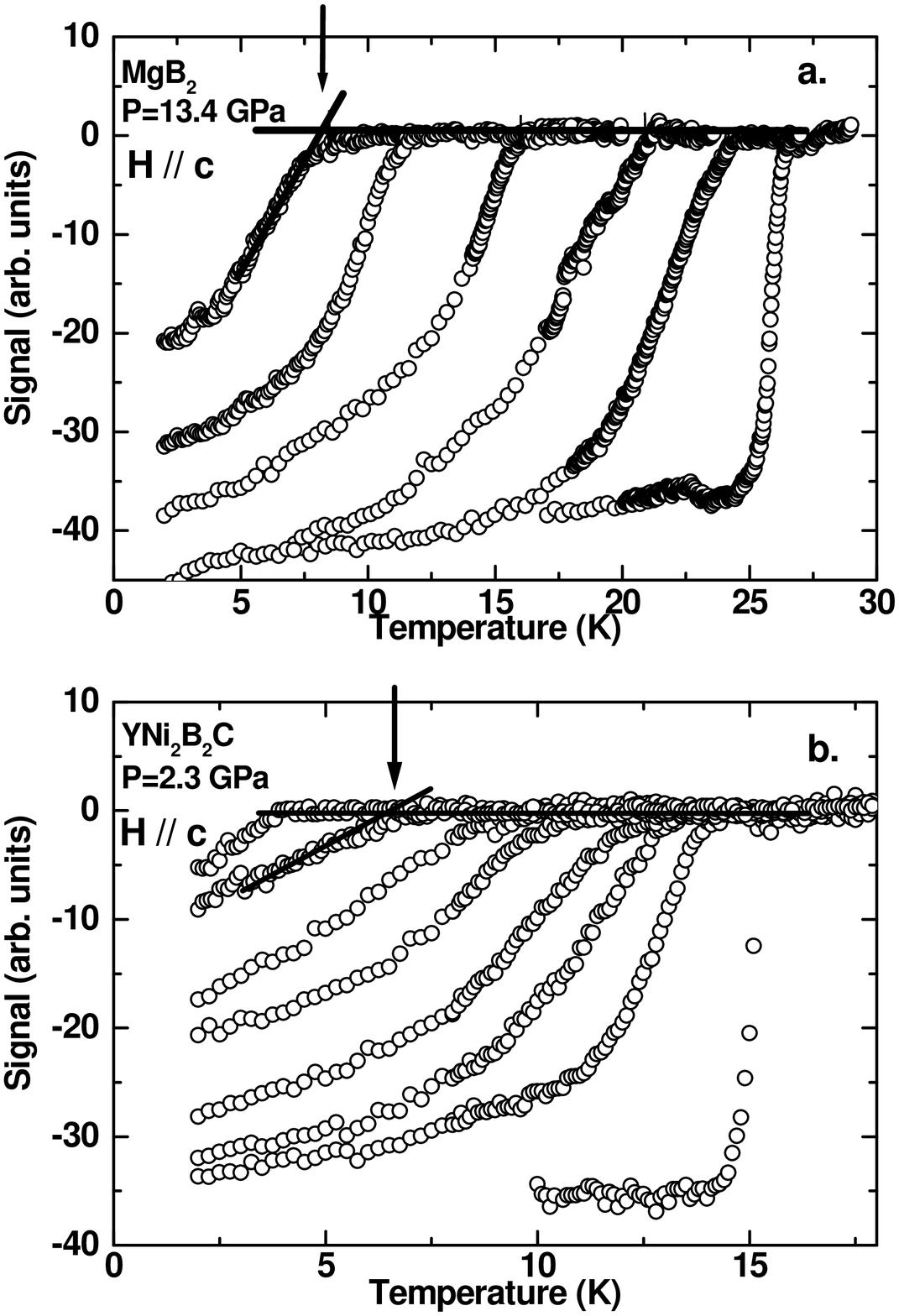}
\caption{The temperature dependence of the ac susceptibility is
shown for several magnetic fields  at 13.4 GPa in MgB$_2$ (from
right to left in a., 0, 0.1, 0.3, 0.7, 1.0, 1.2 T) and at 2.3GPa
in YNi$_2$B$_2$C (from right to left in b., 0, 0.1, 0.3, 0.5, 1.0,
2.0, 3.0, 5.0 T). Lines and arrows graphically demonstrate the way
we extract the critical temperature (onset criterion, see also
text). This series of curves is a representative example of the
transitions measured to obtain the H$_{c2}(T)$ data shown in the
Figs. 2 (MgB$_2$) and 4 (YNi$_2$B$_2$C).}\label{fig:Figsus}
\end{figure}

\newpage
\begin{figure}[ht]
\includegraphics[width=13cm,clip]{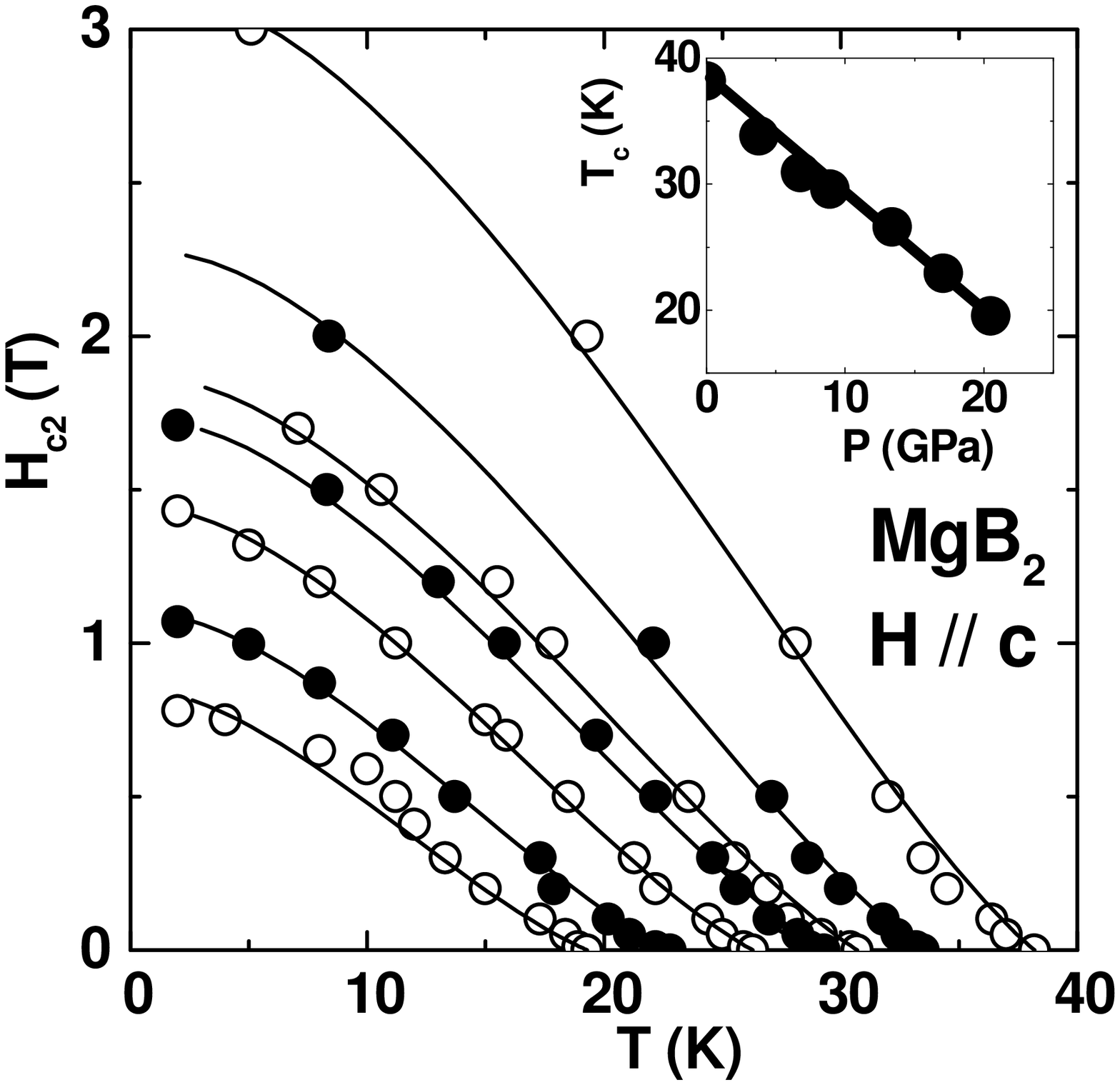}
\vskip -2.5cm \caption{Temperature dependence of the upper
critical field under hydrostatic pressure of MgB$_2$ at, from top
to bottom, ambient pressure, 3.8, 6.8, 8.9, 13.4, 17.1 and 20.5
GPa. The magnetic field was applied along the c axis. The solid
lines are fits discussed in the text. The inset shows the pressure
dependence of the critical temperature obtained here (points). We
obtain dT$_c$/dP=-1.1 K/GPa (line), in agreement with previous
work in high quality single crystalline samples using helium as a
pressure transmitting medium\protect\cite{Deemyad,Goncharov02}.}
\label{fig:Fig1}
\end{figure}

\newpage
\begin{figure}[ht]
\includegraphics[width=13cm,clip]{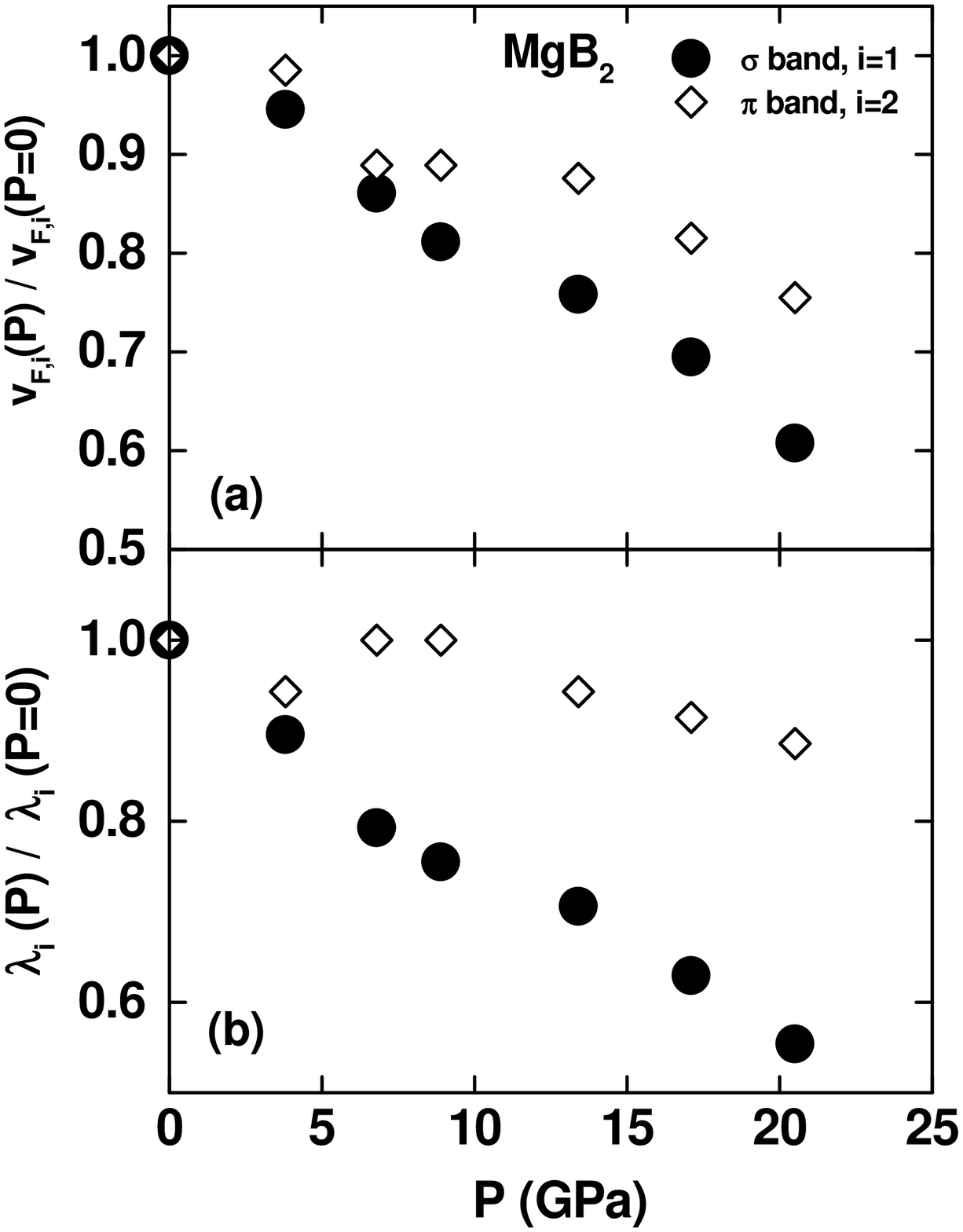}
\vskip -1.5cm \caption{Pressure dependence of the parameters
obtained from the fits of Fig.\ 1 in MgB$_2$, normalized to their
ambient pressure values (Table I). Black points represent the data
for the $\sigma$ bands and open diamonds represent data for the
$\pi$ bands. The electron-phonon coupling parameter $\lambda$ (b)
decreases with pressure, together with the Fermi velocities
$v_{F1,F2}=v_{\sigma,\pi}$ for both bands (a). Note that
$\lambda_{\sigma}$ decreases much faster with pressure than
$\lambda_{\pi}$ (b).}\label{fig:Fig2}
\end{figure}

\newpage
\begin{figure}[ht]
\includegraphics[width=13cm,clip]{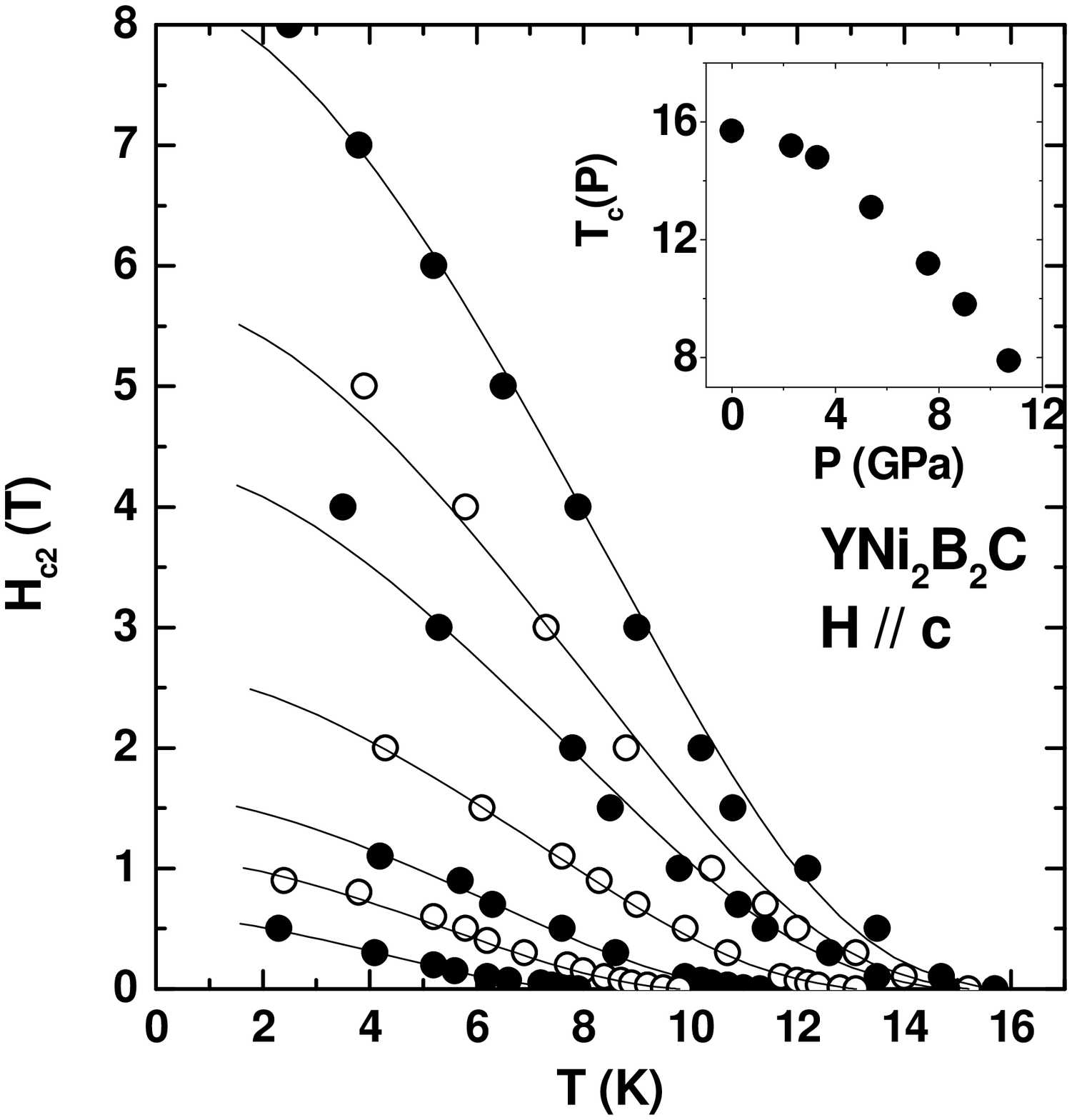}
\vskip -1.5cm \caption{The temperature dependence of the upper
critical field of the Ni borocarbide superconductor YNi$_2$B$_2$C
at, from top to bottom, ambient pressure, 2.3, 3.3, 5.4, 7.6, 9.0
and 11.7 GPa together with the fits to the theory (filled lines)
explained in the text. The inset shows the pressure dependence of
the zero field critical temperature.} \label{fig:Fig3}
\end{figure}

\newpage
\begin{figure}[ht]
\includegraphics[width=13cm,clip]{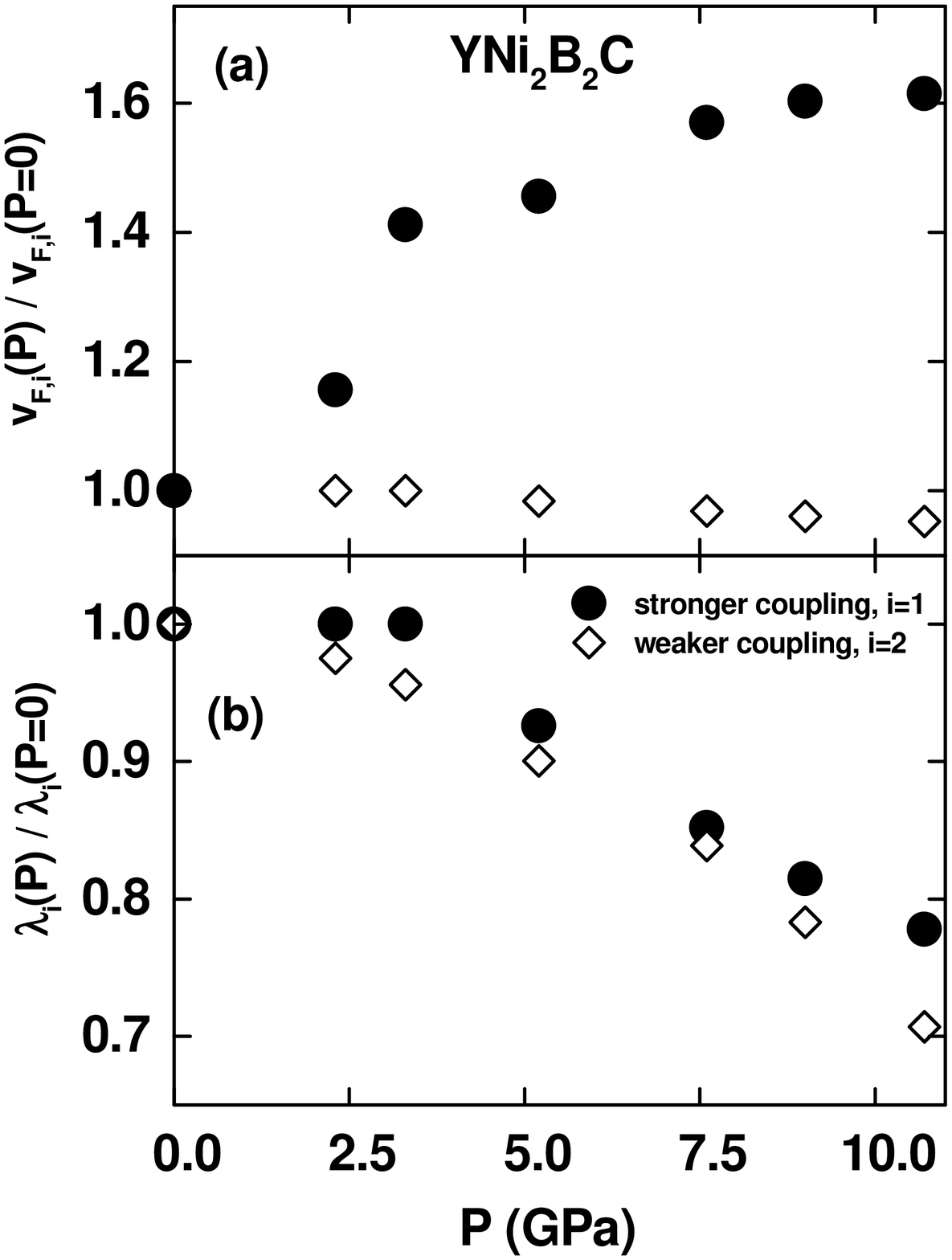}
\vskip -1.5cm \caption{The pressure dependence of the Fermi
surface velocities (a) and  electron-phonon coupling parameters
(b) for both subgroup of electrons is shown as a function of
pressure in YNi$_2$B$_2$C. They are normalized to their ambient
pressure values (see Table I). The subgroup of electrons
corresponding to the strongest coupling part of the Fermi surface,
increases its Fermi velocity $v_{F1}$ (a), and the electron-phonon
coupling (b) is weakened under pressure.} \label{fig:Fig4}
\end{figure}

\end{document}